# Mechanotunable monatomic metal structures at graphene edges

Ning Wei[a,b], Cheng Chang[a,b], Hongwei Zhu[b,c,*] and Zhiping Xu[a,b,*]

[a]Applied Mechanics Laboratory, Department of Engineering Mechanics, Tsinghua University, Beijing 100084, China

[b]Center for Nano and Micro Mechanics, Tsinghua University, Beijing 100084, China

[c]School of Materials Science and Engineering, Key Laboratory of Materials Processing Technology, Tsinghua University, Beijing 100084, China

[*]Authors to whom correspondence should be addressed.

Emails: xuzp@tsinghua.edu.cn (Z.X), hongweizhu@tsinghua.edu.cn (H. Z.)

## Abstract

Monatomic metal (e.g. silver) structures could form preferably at graphene edges. We explore their structural and electronic properties by performing density functional theory based first-principles calculations. The results show that cohesion between metal atoms, as well as electronic coupling between metal atoms and graphene edges offer remarkable structural stability of the hybrid. We find that the outstanding mechanical properties of graphene allow tunable properties of the metal monatomic structures by straining the structure. The concept is extended to metal rings and helices that form at open ends of carbon nanotubes and edges of twisted graphene ribbons. These findings demostrate the role of graphene edges as an efficient one-dimensional template for low-dimensional metal structures that are mechanotunable.

**Introduction**

Monatomic metal structures such as metal chains, rings and helices have attracted notable attention for many years due to their roles as model systems for low-dimensional material studies, as well as their potential in promising applications such as quantum electronics.[1-4] The one-dimensional (1D) nature of these structures is the origin of a number of distinctive behaviors that are absent in their two-dimensional (2D, i.e. mono- or multi-layered) and bulk counterparts.[5] For example, the Fermi-liquid picture for electrons breaks down spectacularly in 1D metals by displaying collective excitations involving spin and charge.[2] On the other hand, electron-phonon interaction and Peierls distortion break lattice symmetry and thus modify the electronic structures of 1D materials.[6] Experimentally, monatomic metal structures can be synthesized by mechanical cleavage of metal nanowires, or metal adatoms assembly epitaxially on the substrate.[5, 7] However, low-dimensional metal structures synthesized by these techniques either lack of considerable structural stability in ambient conditions,[1] or suffer from strongly electronic coupling with the substrate that substantially breaks their 1D nature.[4] Alternative approaches to fabricate stable and well-controlled monatomic metals are thus desired for the research and applications of such a novel structure.

Graphene is another monatomic material that extends in 2D, featuring outstanding structural, mechanical, thermal stabilities and tunable electronic structures.[8-10] There are emerging interests in using graphene as templates to prepare low-dimensional materials and structures.[11] Their tailored nanostructures such as nanoribbons and islands with open edges offer a new dimension following this templating concept for adhesive attachment and self-organization of guest atoms or molecules.[12-14] In a recent experimental work, we reported in-situ synthesis of graphene-metal hybrids where silver nanoplates were synthesized through the templating effect of graphene.[15] The results suggest that metal adheres preferentially to defected sites or edges of graphene.[15] This experimental observation gives us a hint of using the graphene edge as a template to grow low-dimensional metal structures. We assess this approach here by performing first-principles calculations. We explore the structures and properties of several monatomic forms of metal, using Ag as an example, and analysis their mechanical and electronic coupling. The calculation results demostrate an efficient template function of graphene edges, and lay the ground for future exploration on low-dimensional materials.

**Methods**

In this work, we perform plane-wave-basis-set-based density functional theory (DFT) calculations. We use the Quantum-Espresso code.[16] Both local density approximation (LDA)[17] and generalized gradient approximation (GGA)[18] for the exchange-correlation functional are used. GGA is applied in our following discussions if the use of LDA is not specified. We impose ultrasoft pseudopotentials for the ion-electron interactions.[19] Energy cut-offs of 37 and 370 Rydberg are set for the plane-wave basis set and charge density grid, respectively. A Monkhorst-Pack grid with 12 *k*-points is used along the periodic direction for Brillouin zone integration. These settings are verified to meet energy convergence below 1 meV per atom. To relax atomic structures, we use the conjugated gradient (CG) algorithm to minimize the total energy, with convergence criteria for force on atom as 0.01 eVÅ$^{-1}$. In variable-cell relaxation for the periodic direction along graphene edge, the residue stress is controlled to be below 0.01 GPa. For the clarity of presentation, we consider spin-polarization for band structure calculation only as its energy difference with unpolarized calculations is much lower compared to other energy terms (**Table 1** and **2**), and has negligible effect on our discussions on the structural stability and strain effects, as verified by our comparative calculations.

**Results and Discussions**

**Atomic structures and binding energies**

To explore the mechanism of metal-graphene hybridization, we assign silver atoms along both zigzag and armchair graphene edges of a graphene nanoribbon (GNR) and optimize their geometries under a stress-free condition along its axis. We first place metal atoms with a maximum line density along the two edges of a GNR, corresponding to a lattice constant for the Ag chain as $a_{Ag} = a_Z$ or $a_A$, where $a_Z = 2.46$ Å and $a_A = 4.26$ Å are the lattice constants for graphene along zigzag and armchair directions, respectively. Equilibrated structure for the 1D isolated Ag monatomic structure features a zigzag pattern in this case,[3] and the interatomic distance is calculated in this work to be $a_{Ag} = 2.67$ Å. However, here we first use a single unit cell of graphene in the calculations and thus the Ag atoms take a linear chain configuration. The effect of zigzag patter and density-dependence of metal binding will be discussed later. We now assess the stabilities of metal binding through the binding energy $E_b$ at different sites, i.e. at the edges (e.g. tips, valleys), or adatom positions on top of the graphene lattice (see annotations in **Figure 1**). Here $E_b$ is defined as the energy difference between the hybrid $E_{G+nM}$ and the summation of the pristine GNR $E_G$ and the isolated metal atom $E_M$. That is, $E_b = (E_G + nE_M - E_{G+nM})/n$, where $n$ is the number of metal atoms bound to the GNR. The calculation results for zigzag graphene nanoribbons (ZGNRs) with $w = 8$ and armchair graphene nanoribbons (AGNRs) with $w = 12, 13$, and 14 are summarized in **Figure 1**, where the width $w$ is defined as the number of carbon chains along the width direction, following the convention used in Ref. [20]. The results show that metal atoms prefer valley sites, and $E_b$ for adatom sites are lower when the site is closer to graphene edge.

With metal binding to its edge, the GNR has negligible out-of-plane distortion due to its well-preserved aromatic nature. The carbon-carbon bond length at the edge changes slightly (e.g. from 0.138 to 0.141 nm for ZGNR), resulting from the charge transfer between Ag and C atoms and interaction between Ag atoms. In the hybrid, the interatomic distance between metal atoms are defined by the lattice constant of graphene. It should also be noted that the energy difference between metal binding at sites 1, 2 and 3 (as annotated in **Figure 1**) for the Ag-ZGNR hybrid are ~0.1 eV from both LDA and GGA calculations. This amplitude of energy difference is not significantly high compared to thermal energy ($k_BT$ ~26 meV at ambient conditions) and thus stability of the monatomic nature may not be distinguished at elevated ambient temperature. Thicker metal decoration, such as clusters or nanowires, may form with thermal fluctuation, substrate perturbation, or at a high density of metal atoms near the edges.

The formation of metal-graphene hybrids can be discussed along two reaction paths, noted as paths **a** and **b** in this work. Each path includes a two-step process. Along path **a**, metal atoms first assemble into an isolated one-dimensional metal chain. Then the chain attaches to the graphene edge and the hybrid forms. While along path **b**, metal atoms attach to the graphene edges one by one first, and then form a continuous metal chain at the edge. In path **b**, we neglect the effect of neighboring metal atoms when calculating the binding energy for a metal atom at the graphene edge by using a four-period supercell along the ribbon direction. The formation energies can thus be defined through two terms for each path, i.e. $E_{a1} = E_{MC}/N - E_M$, $E_{a2} = E_{MCG}/N - (E_{MC}/N + E_G/N)$, $E_{b1} = E_{MG} - (E_M + E_G/N)$, $E_{b2} = E_{MCG}/N - E_{MG}$, where $N$ is the number of unit cells for the metal-graphene hybrid, and subscripts M, MC, MG, MCG stand for the metal atom, metal chain, single metal atom attachment to the edge, and the hybrid consisting of a metal chain attached to the grapehene edge. Our DFT calculations show that for Ag-ZGNR with $w = 8$, $E_{a1} = -1.21$ eV, $E_{a2} = -1.66$ eV, $E_{b1} = -3.48$ eV, $E_{b2} = 0.41$ eV. These results suggest that along path **a**, two steps are both exothermic, while along path **b**, an energy penalty is required to densify the monatomic metal structure along graphene edge due to their interaction. Clusters with distance larger than the lattice constant $a_{Ag}$ may thus form preferentially.

More specifically for path **b**, where the metal monatomic structures grow on the edge, we further consider the initial steps of metal binding. After an Ag atom is attached to the zigzag edge of graphene, the second atom could attach at the nearest-neighbor or the second nearest-neighbor site. These two sites have binding energies of 3.32 and 3.41 eV respectively (for the second Ag atom), lower than the value for the first attachment (3.48 eV). The nearest binding requires 0.1 eV more energy due to the repulsive force between Ag atoms. Thus the growth of Ag monatomic structures at graphene edges could start by forming low-density patterns and then being desified. The third atom could thus attach to the position along the edge, or on top of the pre-existing Ag dimer and form a trimer. Our calculations show that the edge position is 0.77 eV more preferred than the top site. Moreover, the activation barrier for the metal atom to diffuse from the trimer position to the edge is calculated to be ~0.5 eV, suggesting that edge diffusion may play an important role in forming low-dimensional structures.

Earlier studies on Ag monatomic chains indicated that zigzag configuration forms in isolated Ag chains or Ag chains bound at zigzag graphene edges.[3, 21] In order to include the alternation in positions of metal atoms along graphene edge, we further use a double-period supercell for ZGNR. Random displacement perturbation is added to the Ag atoms in all three directions with amplitude of 0.1 Å to break the mirror symmetry along the graphene edge. From the calculation results, we identify two types of Ag configurations aligned to graphene edges, i.e. at the tip or valley sites along the edge. Both configurations feature a zigzag pattern with respect to the graphene basal plane. As the equilibrium Ag-Ag distance in a metal chain (2.67 Å) is larger than the length of a ZGNR unit cell (2.46 Å), the metal chain thus prefers zigzag structures after it binds to the graphene edge. This can also be seen from the structural information of Ag-ZGNR hybrids listed in **Table 1**, which shows an Ag-Ag distance of 2.67 Å in the zigzag chain, and Ag-C distance of 2.35 Å. The binding energies $E_b$, i.e. 2.82 and 2.87 eV for the tip and valley (**Figure 1e**) configurations, are more than two times higher than $E_b$ for Ag in an isolated chain with the same configuration (1.20-1.21 eV). These results further indicate that interaction between Ag atoms contributes significantly to structural stability of the Ag-ZGNR hybrid.

For AGNRs, as the interatomic distance between neighboring Ag atoms (4.26 Å) is larger than the value in the Ag chain (2.67 Å), no out-of-plane displacement of the Ag atoms is observed due to the weakened interaction between Ag atoms along the graphene edge. The binding of metal atoms to both ZGNR and AGNR edges features apparent density dependence due to the interatomic coupling between Ag atoms. Our first-principles calculations show that an isolated zigzag Ag chain is stable at the interatomic distance $d$ = 2.67 Å. However, a structural transition to the linear configuration occurs with $d$ = 2.78 Å after the Ag chain is elongated, which eventually fails at $d$ = 3.0 Å. These results further confirms our observation that the interatomic interaction between Ag is repulsive for Ag-ZGNR at $a_{Ag} = a_Z$, and almost zero for Ag-AGNR $a_{Ag} = a_A$. One would thus expect for Ag-ZGNR with $a_{Ag} = 2a_Z$, the Ag-Ag interaction will be negligible and a linear configuration will be more favorable, at a reduced density. This is verified by our DFT calculations.

**Electronic coupling**

We now turn to discuss the electronic coupling between metal atoms and the graphene edge. Bader atomic charge analysis[22] suggests a 0.25$e$ charge transfer from Ag to carbon at valley sites along ZGNR edge for $w$ = 8. While for AGNRs, the amount of charge transfer is 0.34$e$ for $w$ = 12, 13, 14. The results also indicate that for both AGNRs and ZGNRs, the metal atoms binding favors valley sites at graphene edges where carbon atoms have two dangling bonds. In contrast, the charge transfer from Ag atoms on top of graphene is much lower, i.e. 0.029$e$ and 0.022$e$ for ZGNR and AGNR (at sites 5 and 6 in **Figure 1a** and **1b**). To further characterize the bonding nature between metal and graphene, distribution of the electron localization function (ELF)[23] is calculated. ELF is a localized function of the ground-state electron density and wavefunction obtained from first-principles calculations. The value of ELF ranges from 0 to 1, where 1 corresponds to the perfect localization as in covalent bonds, and 0.5 corresponds to the electron-gas-like pair probability as in metallic bonds. The results in **Figure 2** show that high ELF regions are localized on Ag atoms, and thus suggest an ionic nature of the Ag-GNR interaction.

These results, in combination with previous discussion on the binding energies, clearly suggest that in addition to the cohesion between metal atoms, the electronic coupling between metal atoms and graphene edge could be another driving force to stabilize the hybrid structures.

To explore the perturbation of metal binding to the electronic structures of both metal and graphene, we analyze band structures and density of states (DOS) obtained from spin-polarized DFT calculations. We consider hybrid structures with metal atoms bound to both sides of the ZGNR in a double-period supercell, i.e. the zigzag pattern of Ag chain forms. Antiferromagnetic (AFM), ferromagnetic (FM) and paramagnetic (PM) configurations are identified. The difference in energy is 51.9 meV per unit cell lower for both AFM and FM than PM ordering, and the difference between AFM and FM states is below 1 meV. The band structures plotted in **Figure 3** show that the electronic structure of Ag monatomic structures remains almost intact, suggesting weakly electronic coupling between Ag atoms and ZGNR. Specifically, charge transfer upshifts bands for electrons in Ag and downshifts those for electrons in graphene, i.e. the graphene edges is doped in *n*-type, and the localized states attributed to the dangling bonds at graphene edges are removed by Ag-termination.

**Strain effects**

One of the signature characteristics of graphene is its exceptional mechanical stability, that enables strain engineering with tensile strain up ~ 20%. Thus by deforming GNRs we could tune the monatomic metal structures accordingly, which is not allowed in monatomic metal wires forming on metal substrates.[1-4] To illustrate the structural evolution of the Ag-ZGNR hybrid (with Ag in the valley position) as a function of axial strain along GNR, we perform tensile tests and monitor changes in the atomic structure. As shown in **Figure 4a**, the out-of-plane displacement of Ag atoms relative to the graphene basal plane surface decreases with strain and becomes zero at 11% strain.

Changes in the Ag-Ag and Ag-C distances depicted in **Figure 4b** shows that before the Ag monatomic structure is flattened to the graphene basal plane, the Ag-Ag distance remains the same, and the structure evolves only by changing the angle between them. In contrast, the graphene is strained affinely and thus defines the mechanical response of the hybrid before it breaks down with fracture nucleates inside the graphene lattice by breaksing $sp^2$ C-C bonds. Accordingly, mechanical properties of the hybrid (stiffness, strength, strain to failure, etc.) are comparable to graphene. Our calculations verify this by demonstrating the strain to failure as 24%, which is the same as the value for graphene and higher than the value 21.74% for the isolated Ag chain. Thus the monatomic Ag structure formed could be modulated from a zigzag configuration to a linear chain. After the tensile strain exceeds 11%, the interatomic distance can be further tuned continuously within the range from 2.67 to 3.06 Å. This mechanotunable function holds great promises in exploring, for example, quantum transport in low-dimensional metal structures and its correlation with atomic structures as well as vibrational effects.

**Metal rings and other monatomic structures**

With the results obtained above for (straight, zigzag) monatomic metal structures formed at graphene edges, more low-dimensional structures could be envisioned. For example, two illustrative examples are explored here, including metal rings formed at ends of carbon nanotubes, and metal helices formed at edges of twisted GNRs.[24, 25] The results show that both these structures are comparably stable with Ag atoms aligned at graphene edges. The structures and electronic density of states (DOS) for Ag-CNT hybrid are summarized in **Table 2** and **Figure 5**, which is consistent with the picture of weak coupling between Ag and CNT ends. Moreover, an alternative pattern of radial displacement (type II) is preferred for small-diameter CNTs (6, 6) and (8, 8), while for (10, 10) CNT with a larger diameter, configuration (type I) with all Ag atoms residing outside of the CNT is preferred. The concept of templating low-dimensional metal structures using graphene edges could be further extended, for example, in forming thin nanowires of metals at the edges of bilayer or multilayer graphene sheets that are staggered, or mono- and multilayers on the basal plane of graphene sheets.[9, 26] It is worth noting that the binding energy of Ag at ZGNR edges is 2.82 eV, that is even higher than Ag in the close-packing 2D triangular lattice (1.92 eV) and 3D face-centered-cubic (FCC) lattice (2.62 eV). Thus these metal nanostructures formed at graphene edges features significant structural stabilities, as demonstrated in our previous experiments.[15]

Although main discussion in this paper is based on Ag hybrid, we also explore binding of Au and Ni atoms at graphene edges that share the common feature of preferred edge binding as we have identified for Ag. These hybrids are interesting as Au nanowire assembled at graphene edges is expected to be more stable than Ag due to its resistance to oxidation, while the presence of *d*-electrons in Ni implies interesting magnetic behaviors.[3] Our calculations show that the formation of monatomic Au and Ni structures at graphene edges are also energy favorable.

The bare edges of graphene are not stable in ambient condition and could be terminated by chemical groups such as hydrogen. We thus probe the effect of hydrogen termination (H-term) by performing DFT calculations with metal atoms placed at the tip and valley positions from the hydrogen atoms. The results summarized in **Table 1** suggest that although the coupling between Ag atoms and graphene edge is weakened, the binding energy 1.24 eV, however, is still significant enough to provide considerable stability at ambient conditions. The value is also close to the binding energy in the isolated Ag chain with the same configuration (1.21 eV). This result thus further validates our conclusion that as the cohesion between metal atoms and electronic coupling between metal and graphene are two key driving forces for the assembly of metal monatomic structures along graphene edges.

## Conclusion

To conclude, our first-principles calculations have revealed that metallic decoration of graphene nanostructures is favored at their edges, and the stability is attributed to both cohesion between metal atoms and metal-graphene interaction. The weak electrostatic coupling between metal atoms and graphene, the outstanding structural stability and mechanical properties of graphene enable a robust templating approach with strain tunability, offering opportunities in low-dimensional material design and applications.

## Acknowledgments

This work was supported by the National Natural Science Foundation of China through Grant 11222217, 11002079, 51372133, and Tsinghua University Initiative Scientific Research Program 2011Z02174, 2012Z02102. This work was performed on the Explorer 100 cluster system of Tsinghua National Laboratory for Information Science and Technology.

# Tables, Figures and Figure Captions

**Table 1.** Structure and energy information for the metal-ZGNR hybrids, including metal-metal and metal-carbon distances $d_{M-M}$ and $d_{M-C}$, the binding energies $E_b$ for Ag attached to graphene edges, and binding energies $E_{chain}$ of Ag in a monatomic chain with the same configuration as in the Ag-ZGNR hybrid.

| structure | binding site | $d_{M-M}$ (Å) | $d_{M-C}$ (Å) | $E_b$ (eV) | $E_{chain}$ (eV) |
|---|---|---|---|---|---|
| Ag-ZGNR | tip | 2.74 | 2.15 | 2.82 | 1.20 |
| | valley | 2.67 | 2.35 | 2.87 | 1.21 |
| Ag-H-ZGNR | tip | 2.67 | 4.26 | 1.24 | 1.21 |
| | valley | 2.67 | 4.54 | 1.24 | 1.21 |

**Table 2.** Structure and energy information for the Ag-CNT hybrid. Atomic structures of type I and II configurations are defined in **Figure 5**, where Ag atoms have the same or alternative radial distance from the carbon atoms in CNT. For type II, two different metal-carbon distances are both listed.

| structure | type | $d_{\text{M-M}}$ (Å) | $d_{\text{M-C}}$ (Å) | $E_b$ (eV) | $E_{\text{chain}}$ (eV) |
|---|---|---|---|---|---|
| Ag-(6,6) CNT | Type I | 2.65 | 2.39 | 2.64 | 1.18 |
| | Type II | 2.70 | 2.28/2.54 | 2.74 | 1.40 |
| Ag-(8,8) CNT | Type I | 2.65 | 2.37 | 2.71 | 1.11 |
| | Type II | 2.75 | 2.35/2.42 | 2.88 | 1.44 |
| Ag-(10,10) CNT | Type I | 2.65 | 2.37 | 2.77 | 1.17 |
| | Type II | 2.68 | 2.32/2.46 | 2.72 | 1.28 |

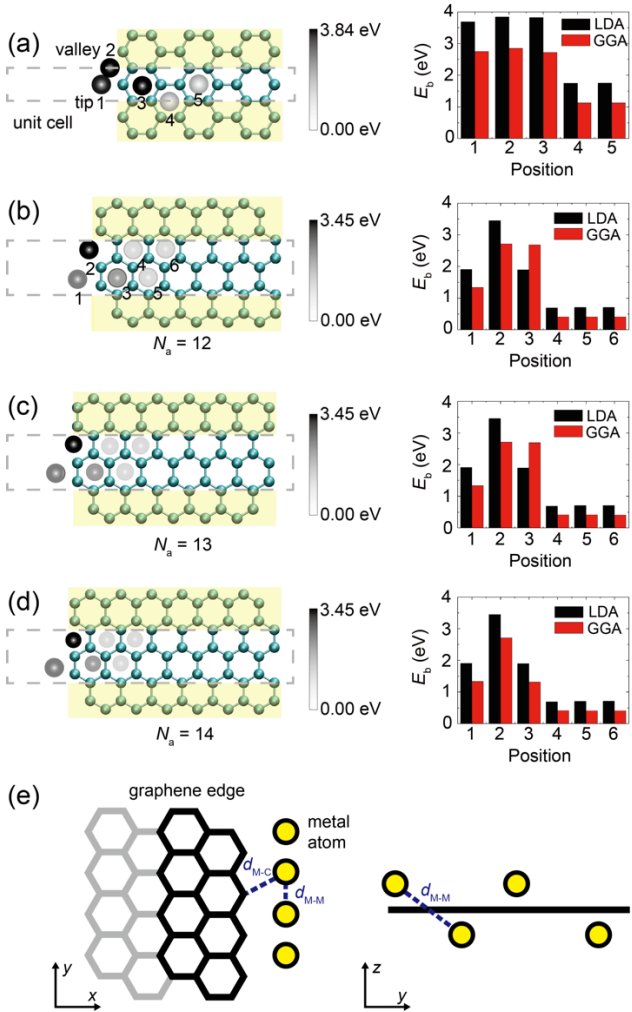

**Fig. 1.** Atomic structures and binding energies of metal atoms bound to graphene edges as a linear chain. (a-d) Binding sites of Ag atoms at ZGNR and AGNR edges, and their relative binding energies dictated by the gray level. (e) Zigzag metal monatomic structures formed at the valley sites of ZGNR edges. $d_{M-C}$ and $d_{M-M}$ are the metal-carbon and metal-metal distances in the hybrid mentioned in the text.

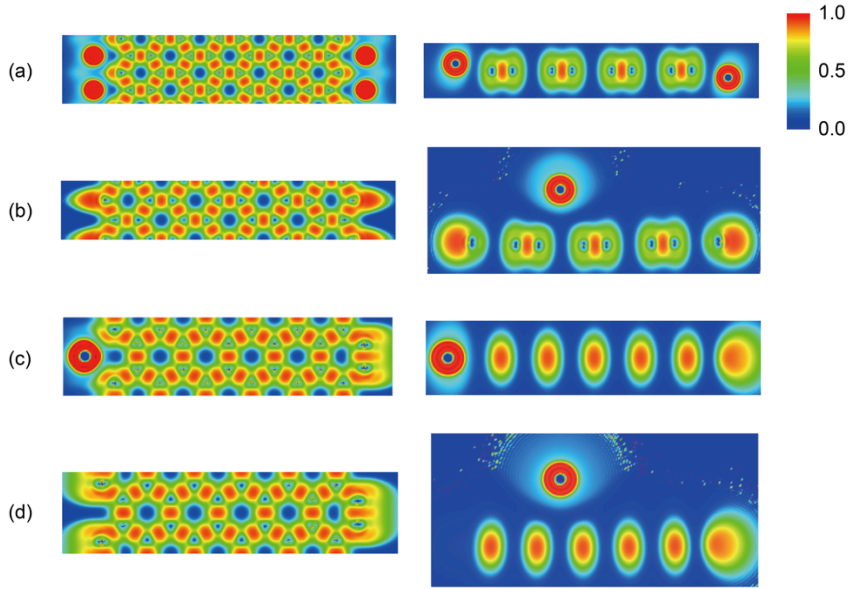

**Fig. 2.** Electron localized function distribution of (a, b) Ag-ZGNR and (c, d) Ag-AGNRs hybrids, with Ag atoms bound to the valley site at graphene edges (a, c) and on top of graphene (b, d). Both top view (left) and side view (right) are plotted.

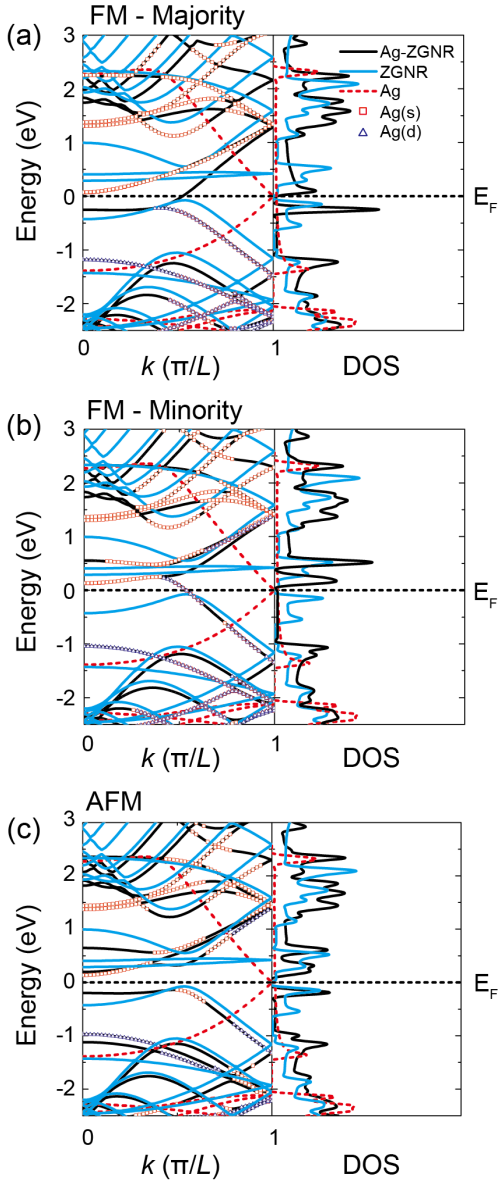

**Fig. 3.** Energy band structures (left) and density of state (DOS, right) for the Ag-ZGNR hybrid, zigzag Ag-chain and bare ZGNR, respectively. The projections on *s*- and *d*-orbitals of Ag atoms in the Ag-ZGNR are plotted using squares and triangles. (a) and (b) are for major and minor spin components in the FM state and (c) is for the AFM state.

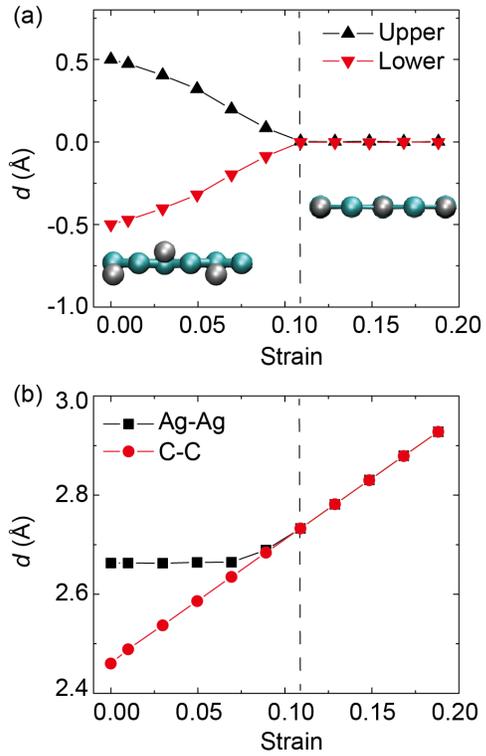

**Fig. 4.** (a) Out-of-plane displacement of Ag atoms measured a function of tensile strain on the double-period zigzag Ag-ZGNR hybrid with Ag at the valley position. (b) Distance between Ag-Ag atoms and C-C atoms at graphene edges, which becomes synchronized at 11% strain.

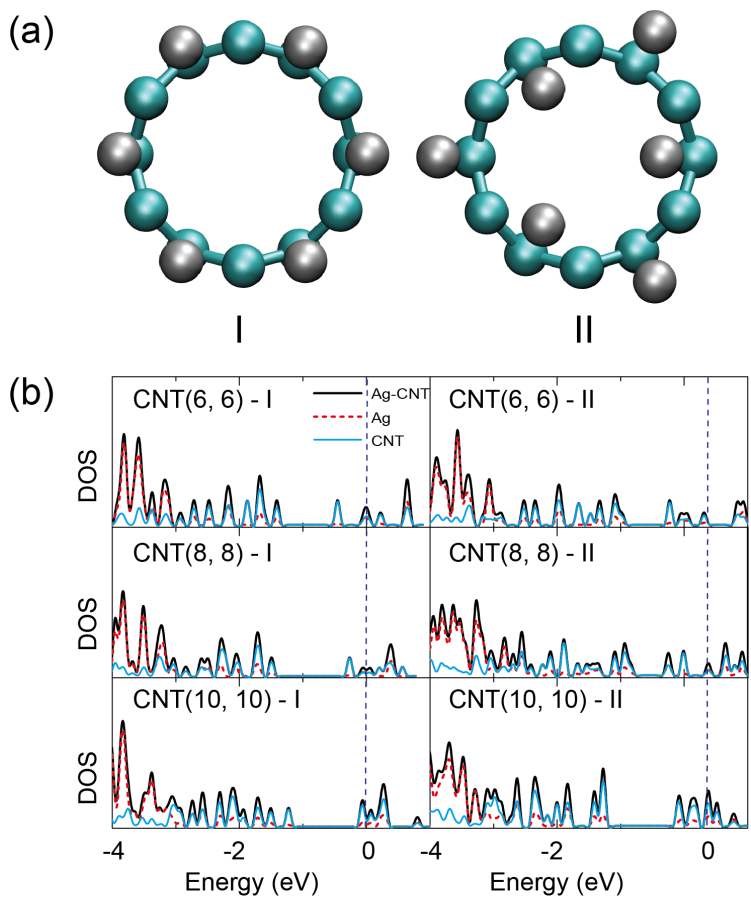

**Fig. 5.** Atomic structures (a) and DOS (b) for the Ag-CNT hybrids. In type I and II configurations, the Ag atoms have the same and alternative radial displacement with respect to carbon atoms in the CNT. The relative stabilities and geometry are summarized in **Table 2**.